\documentclass[conference]{IEEEtran}
\usepackage{graphicx,amsmath,latexsym,amssymb,subfigure}
\usepackage[table,xcdraw]{xcolor}
\begin{document}
%
\title{Are mmWave Low-Complexity 
Beamforming Structures Energy-Efficient? Analysis of the Downlink MU-MIMO}
%
%
%

\author{\IEEEauthorblockN{Stefano Buzzi  and Carmen D'Andrea}
\IEEEauthorblockA{DIEI - Universit\`a di Cassino  e del Lazio Meridionale\\
I-03043 Cassino (FR) - Italy \\
{\tt \{buzzi, carmen.dandrea\}@unicas.it}}
}
\maketitle

\begin{abstract}
Future cellular systems based on the use of above-6 GHz frequencies, the so-called millimeter wave (mmWave) bandwidths, will heavily rely on the use of antenna arrays both at the transmitter and at the receiver, possibly with a large number of elements. For complexity reasons, fully digital precoding and postcoding structures may turn out to be unfeasible, and thus suboptimal structures, making use of simplified hardware and a limited number of RF chains, have been investigated. This paper considers and makes a comparative assessment, both from a spectral efficiency and energy efficiency point of view, of several suboptimal precoding and postcoding beamforming structures for the downlink of a cellular multiuser MIMO (MU-MIMO) system. 
Based on the most recently available data for the energy consumption of phase shifters and switches, we show that there are cases where fully-digital beamformers may achieve a larger energy efficiency than lower-complexity solutions, as well as that structures based on the exclusive use of switches achieve quite unsatisfactory performance in realistic scenarios. 
\end{abstract}


\section{Introduction}
The use of frequency bands in the range $10-100$ GHz, a.k.a. millimeter waves (mmWaves), for cellular communications, is among the most striking technological innovations brought by  fifth generation (5G)  wireless networks \cite{Buzzi5G}.  The scarcity of available frequency bands in the sub-6 GHz spectrum has been the main thrust for considering the use of higher frequencies for cellular applications, and indeed recent research \cite{rappaport2013millimeter,HeathBook} has shown that mmWaves, despite increased path-loss and atmospheric absorption phenomena, can be actually used for cellular communications over short-range distances (up to 100-200 meters), provided that multiple antennas are used at both sides of the communication link: MIMO processing, thus, is one distinguishing and key feature of mmWave systems. 


When considering MIMO architectures, hardware complexity and energy consumption issues make the use of conventional fully digital beamforming, which requires one RF chain for each antenna element, rather prohibitive; 
as a consequence, recent research efforts have been devoted towards devising suboptimal, lower complexity, beamforming structures \cite{chih-lin-i}. In particular, hybrid beamforming structures have been proposed, with 
a limited number (much smaller than the number of antenna elements) of RF chains. The paper \cite{alkhateeb2015limited}  analyzes the achievable rates for a MU-MIMO system with hybrid precoding and limited feedback; it is therein shown that, for the case of single-path (i.e., rank-1) channels hybrid precoding structures achieve a spectral efficiency very close to those of a fully digital beamformer. In \cite{sohrabi2016hybrid}, it is shown that a hybrid beamformer with a number of RF chains that is twice the number of transmitted data streams may exactly mimic a fully digital beamformer; the analysis, which neglects energy efficiency issues,  is  however limited to either a single-user MIMO system or a MU-MIMO system with single-antenna receivers. The paper 
\cite{SwitchesRial} proposes a new low-complexity combining structure, based on switches rather than on analog phase shifters; the performance of this new structure is evaluated in a rather simple scenario, i.e. single-user MIMO system with a limited number of transmit and receive antennas. The paper \cite{Switches_constantPS} although relative to sub-6 GHz frequencies, introduces a novel combining structure made of fixed (rather than tunable) phase shifters and of switches; this paper  focuses on the receiver (equipped with a large array), while the transmitters have only one antenna. In \cite{MIMOArchitecture}, the authors considers five different low-complexity decoding structures, all based on the use of phase shifters and switches, and provide  an analysis of the achievable spectral efficiency along with estimates of the energy consumption of the proposed structures. The paper, however, does not analyze the system energy efficiency (i.e. the ratio of the achievable rate to the consumed power \cite{buzziJSAC2016}), and focuses only on the receiver omitting a similar analysis for the transmitter implementation. In \cite{roth2016channel} a consumed power model for components designed for 60 GHz is given, and a comparison between fully-digital beamforming, 1-bit analog-to-digital conversion, and analog beamforming is given.

This paper, to the best of our knowledge, is the first to focus not only on the achievable spectral efficiency (ASE), but also on the topic of energy efficiency for hybrid low-complexity precoding and combining structures. With reference to a multiuser MIMO (MU-MIMO) system operating at mmWave frequencies, we consider several of these structures, taking into account both the transmitter and the receiver, and provide an analysis of both the ASE and of the global energy efficiency (GEE). While the results on the ASE can be deemed as "stable" and are in agreement with intuition and existing results, things are a little bit more involved when considering the GEE. Indeed, here, the relative ranking of the several low-complexity structures strongly depends on the adopted power consumption model for amplifiers, phase shifters, switches, etc. As an instance, our results show that, using recent power models, there are instances where fully digital beamforming may still be the most energy-efficient solution. 

This paper is organized as follows. Next Section contains the system model; Section III describes the several low-complexity beamforming structures along with the used power consumption models, while Section IV contains the discussion of the obtained numerical results. Concluding remarks are finally given in Section V.

\section{The system model}
We consider the downlink of a single-cell MU-MIMO system wherein one BS communicates, on the same frequency slot, with $K$ mobile users. 
We denote by $N_T$ the number of transmit antennas at the BS, and by $N_R$ the number of receive antennas at the user's device\footnote{For the sake of simplicity we assume that all the mobile receivers have the same number of antennas; however, this hypothesis can be easily relaxed.}. 

\subsection{Channel model}

We focus on a narrowband clustered channel model, 
so that the baseband equivalent of the propagation channel between the transmitter and the generic receiver\footnote{For ease of notation, we omit, for the moment, the subscript $"k"$ to denote the BS to the $k$-th user channel matrix.}
is  represented by an $(N_R \times N_T)$-dimensional matrix expressed as:
\begin{equation}
\mathbf{H}=\gamma\sum_{i=1}^{N_{\rm cl}}\sum_{l=1}^{N_{{\rm ray},i}}\alpha_{i,l}
\sqrt{L(r_{i,l})} \mathbf{a}_r(\phi_{i,l}^r) \mathbf{a}_t^H(\phi_{i,l}^t) + \mathbf{H}_{\rm LOS}\; .
\label{eq:channel1}
\end{equation}
In Eq. \eqref{eq:channel1}, 
we are implicitly assuming that the propagation environment is made of $N_{\rm cl}$ scattering clusters, each of which contributes with $N_{{\rm ray}, i}$ propagation paths, $i=1, \ldots, N_{cl}$, plus a  possibly present LOS component.  
We denote by  $\phi_{i,l}^r$ and $\phi_{i,l}^t$ the angles of arrival and departure of the $l^{th}$ ray in the $i^{th}$ scattering cluster, respectively. 
The quantities $\alpha_{i,l}$ and $L(r_{i,l})$ are the complex path gain and the attenuation associated  to the $(i,l)$-th propagation path. 
The factors $\mathbf{a}_r(\phi_{i,l}^r)$ and $\mathbf{a}_t(\phi_{i,l}^t)$ represent the normalized receive and transmit array response vectors evaluated at the corresponding angles of arrival and departure; for an uniform linear array (ULA) with half-wavelength inter-element spacing we have:
\begin{equation}
\mathbf{a}_t(\phi_{i,l}^t)=\displaystyle \frac{1}{\sqrt{N_T}}[1 \; e^{-j\pi \sin \phi_{i,l}^t} \; \ldots \; e^{-j\pi (N_T-1) \sin \phi_{i,l}^t}] \; ;
\label{eq:ULA}
\end{equation}
A similar expression can be also given for $\mathbf{a}_r(\phi_{i,l}^r)$.
Finally, $\gamma=\displaystyle\sqrt{\frac{N_R N_T}{\sum_{i=1}^{N_{\rm cl}}N_{{\rm ray},i}}}$  is a normalization factor ensuring that the received signal power scales linearly with the product $N_R N_T$. 
A detailed description of all the parameters needed for the generation of sample realizations for the channel model of Eq. \eqref{eq:channel1} is reported in \cite{buzzidandreachannel_model}, and we refer the reader to this reference for further details on the channel model.

\subsection{Transmitter and Receiver Processing}

We denote by $M$ the number of data symbols sent to each user in each signalling interval\footnote{Otherwise stated, the BS transmits in each time-frequency slot $M K$ data symbols.}, and by $\mathbf{x}_k$ the $M$-dimensional vector of the data symbols intended for the $k$-th user; the discrete-time signal transmitted by the BS can be expressed as the $N_T$-dimensional vector
$\mathbf{s}_T=  \sum_{k=1}^{K} \mathbf{Q}_k \mathbf{x}_k$, with $\mathbf{Q}_k$ the $(N_T \times M)$-dimensional precoding matrix for the $k$-th user. The signal received by the generic $k$-th user is expressed as the following $N_R$-dimensional vector
\begin{equation}
\mathbf{y}_k= \mathbf{H}_k \mathbf{s}_T + \mathbf{w}_k \; ,
\label{eq:yk}
\end{equation}
with $\mathbf{H}_k$ representing the clustered channel (modeled as in Eq. \eqref{eq:channel1}) from the BS to the $k$-th user and $\mathbf{w}_k$ is the $N_R$-dimensional additive white Gaussian noise with zero-mean i.i.d. entries with variance $\sigma_n^2$. Denoting by $\mathbf{D}_k$ the $(N_R \times M)$-dimensional post-coding matrix at the $k$-th user device, the following $M$-dimensional vector is finally obtained:
\begin{equation}
\mathbf{r}_k= \mathbf{D}_k^H \mathbf{H}_k \mathbf{Q}_k \mathbf{x}_k + \sum_{\ell =1, \ell \neq k}^K
\mathbf{D}_k^H \mathbf{H}_k \mathbf{Q}_\ell \mathbf{x}_\ell + \mathbf{D}_k^H \mathbf{w}_k  \; .
\end{equation}

Now, depending on the choice of the precoding and postcoding matrices $\mathbf{Q}_k$ and $\mathbf{D}_k$,  several transceiver structures can be conceived. These will be illustrated later in section III.

\subsection{The considered performance measures}

We will consider two performance measures: the ASE and the GEE. 
The ASE is measured in [bit/s/Hz], while the GEE is measured in 
[bit/Joule] \cite{buzziJSAC2016}.
Assuming Gaussian data symbols\footnote{The impact on the ASE of a finite-cardinality modulation is a topic worth future investigations.}, the ASE for the downlink case is well-known to be expressed as\footnote{Note that we are here assuming that the power budget $P_T$ is uniformly divided among the data streams, although power allocation could be easily performed.}
\begin{equation}
{\rm ASE}= \displaystyle \sum_{k=1}^K \log \det \left[ \mathbf{I}_M + \frac{P_T}{KM}\mathbf{R}_{\overline{k}}^{-1}\mathbf{D}_k^H \mathbf{H}_k
\mathbf{Q}_k \mathbf{Q}_k^H \mathbf{H}_k^H \mathbf{D}_k
\right] \; ,
\label{eq:ASE}
\end{equation}
 wherein $\mathbf{I}_M$ is the identity matrix of order $M$, $P_T$ is the BS transmit power, and 
 $\mathbf{R}_{\overline{k}}$ is the covariance matrix of the overall disturbance seen by the $k$-th user receiver, i.e.:
\begin{equation}
\mathbf{R}_{\overline{k}}=\sigma^2_n \mathbf{D}_k^H \mathbf{D}_k +
\frac{P_T}{MK} \sum_{\ell =1, \ell \neq k}^K
\mathbf{D}_k^H \mathbf{H}_k \mathbf{Q}_\ell \mathbf{Q}_\ell^H \mathbf{H}_k^H \mathbf{D}_k \; .
\end{equation}
Regarding the GEE, it is defined as 
\begin{equation}
{\rm GEE}= \displaystyle \frac{ W {\rm ASE}}{\eta P_T + P_{\rm{TX},c}+K P_{\rm{RX},c}} \; ,
\label{eq:GEE}
\end{equation}
where  $W$ is the system bandwidth,  $P_{\rm{TX},c}$ is the amount of power consumed by the BS circuitry,  $P_{\rm{RX},c}$ is the amount of power consumed by the mobile user's device circuitry, and 
$\eta>1$ is a scalar coefficient modelling the power amplifier inefficiency. Note that, differently from what happens in the most part of existing studies on energy efficiency for cellular communications (see, for instance, references of \cite{buzziJSAC2016}), we have included in the GEE definition \eqref{eq:GEE}  the power consumed both at the BS and at the mobile user's devices.  

\section{Beamforming structures}
In the following, we detail the  beamforming pre-coding and post-coding structures considered in this work. 

\subsection{Channel-matched, fully-digital (CM-FD) beamforming}

Letting $\mathbf{H}_k= \mathbf{U}_k \mathbf{\Lambda}_k \mathbf{V}_k^H$ denote the singular-value-decomposition (SVD) of the matrix $\mathbf{H}_k$, the $k$-th user pre-coding and combining matrices $\mathbf{Q}_k^{\rm CM-FD}$ and  $\mathbf{D}_k^{\rm CM-FD}$  are chosen as the columns of the matrices $\mathbf{V}_k$ and 
$\mathbf{U}_k$, respectively, corresponding to the $M$ largest entries in the eigenvalue matrix $\mathbf{\Lambda}_k$. 
The CM-FD beamforming is optimal in the interference-free case, and tends to be optimal in the case in which the number of antennas at the transmitter grows large. The considered fully digital precoding architecture requires a baseband digital precoder that adapts the $M$ data streams  to the $N_T$ transmit antennas; then, for each antenna there is a  digital-to-analog-converter (DAC), an RF chain and a power amplifier (PA). At the receiver,  a low noise amplifier (LNA), an RF chain, an analog-to-digital converter (ADC) is required for each antenna, plus a baseband digital combiner that combines the $N_R$ outputs of ADC to obtain the soft estimate of the $M$ trasmitted symbols. The amount of power consumed by the transmitter circuitry can be thus expressed as:
\begin{equation}
P_{\rm{TX},c}=N_T\left(P_{\rm RFC}+P_{\rm DAC}+P_{\rm PA}\right)+P_{\rm BB} \;  ,
\end{equation} 
and the amount of power consumed by the receiver circuitry can be expressed as:
\begin{equation}
P_{\rm{RX},c}=N_R\left(P_{\rm RFC}+P_{\rm ADC}+P_{\rm LNA}\right)+P_{\rm BB} \;  .
\end{equation} 
In the above equations, $P_{\rm RFC}= 40$ mW  \cite{MIMOArchitecture} is the power consumed by the single RF chain, $P_{\rm DAC}= 110$ mW \cite{DAC_vandenbosh} is the power consumed by each DAC, 
$P_{\rm ADC}=200$mW  \cite{MIMOArchitecture} is the power consumed by each single ADC, 
$P_{\rm PA}=16 $ mW  \cite{PhasedArray60GHz} is the power consumed by the PA,
$P_{\rm LNA}=30$ mW \cite{MIMOArchitecture} is the power consumed by the LNA,  
and $P_{\rm BB}$ is the amount of power consumed by the baseband precoder/combiner designed; assuming a  CMOS implementation we have a power consumption of 243 mW \cite{BBprec_combiner}.

\subsection{Partial zero-forcing, fully digital (PZF-FD) beamforming}
Zero-forcing precoding nulls interference at the receiver through the constraint that the $k$-th user precoder be such that the product $\mathbf{H}_\ell \mathbf{Q}_k$ is zero for all $\ell \neq k$. In order to avoid a too severe noise enhancement, we resort here to a partial zero-forcing, namely we require that the columns of the precoding matrix $\mathbf{Q}_k$ are orthogonal to the $M$ (the number of transmitted data-streams to each user) right eigenvectors of the channel $\mathbf{H}_\ell$ corresponding to the largest eigenvalues of $\mathbf{H}_\ell$, for all $\ell \neq k$. In this way, the precoder orthogonalizes only to a $M(K-1)$-dimensional subspace and nulls the most significant part of the interference.  Formally, the precoder $\mathbf{Q}_k^{\rm PZF-FD}$ is obtained as the projection of the CM-FD precoder $\mathbf{Q}_k^{\rm CM-FD}$ onto the orthogonal complement of the subspace spanned by the $M$ dominant right eigenvectors of the channel matrices $\mathbf{H}_1, \ldots, \mathbf{H}_{k-1}, \mathbf{H}_{k+1}, \ldots, \mathbf{H}_K$. The combining matrix is instead obtained as $\mathbf{D}_k^{\rm PZF-FD}=(\mathbf{H}_k \mathbf{Q}_k^{\rm PZF-FD})^+$, with $(\cdot)^+$ denoting pseudo-inverse. Since the PZF-FD beamforming requires a fully digital combining, its power consumption is the same as that of the CM-FD beamformer.

\subsection{Partial zero-forcing, hybrid (PZF-HY) beamforming}
\begin{figure}
\begin{subfigure}
  \centering
  \includegraphics[scale=0.41]{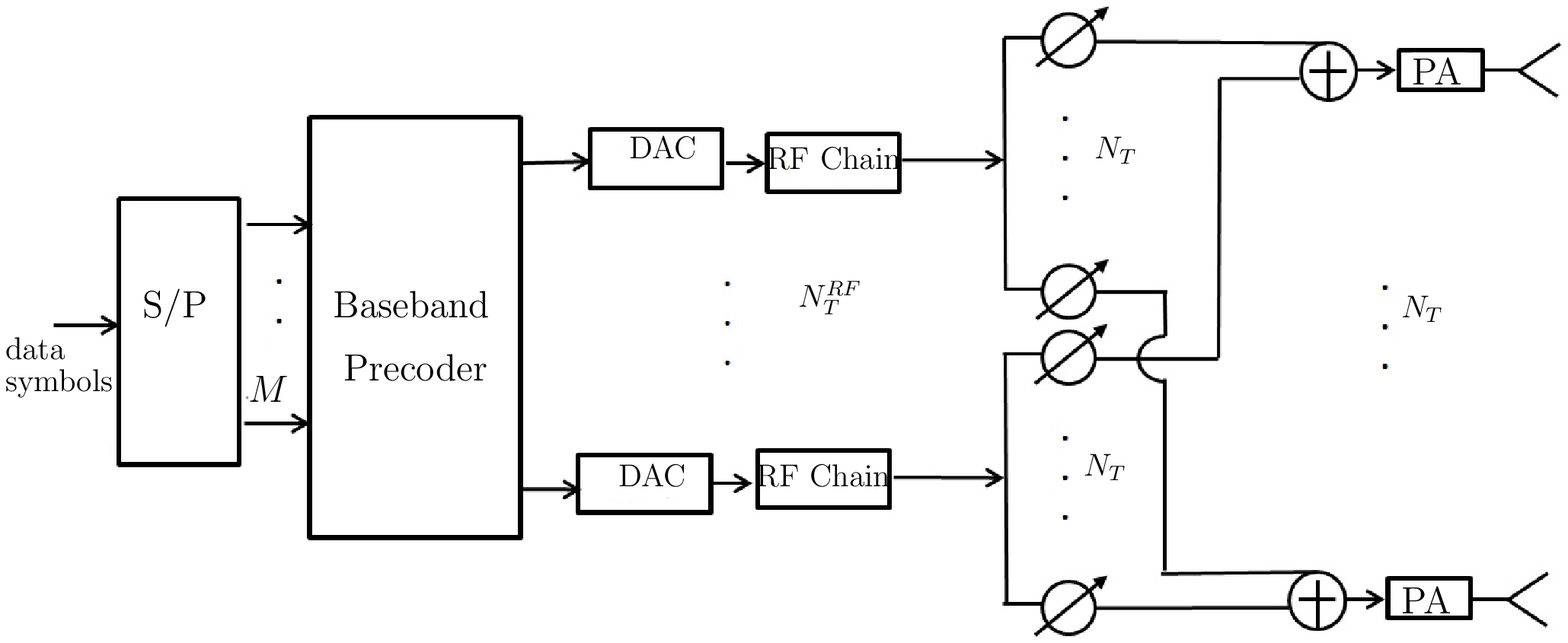}
\end{subfigure}%
\begin{subfigure}
  \centering
  \includegraphics[scale=0.41]{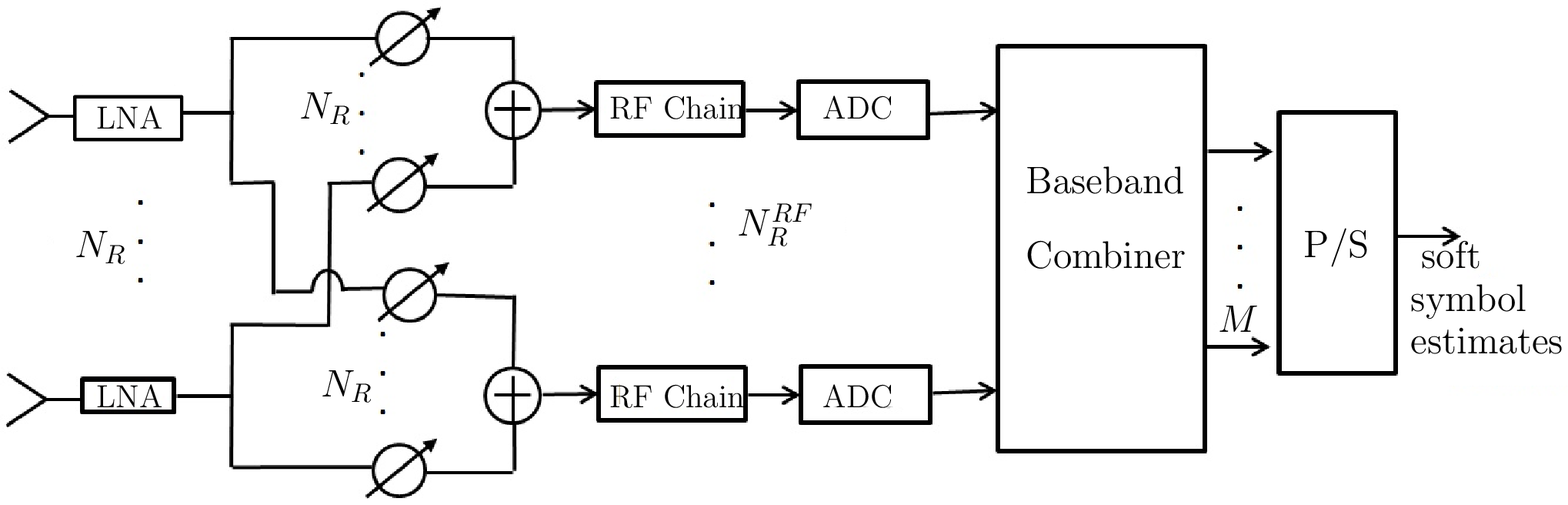}
\end{subfigure} 
\caption{Block-scheme of a transceiver with hybrid digital/analog beamforming.}
\label{Fig:hybrid_structure}
\end{figure}

In order to avoid a number of RF chains equal to the number of antennas, hybrid beamforming architectures have been proposed; in particular, denoting by $N_T^{\rm RF}$ and $N_R^{\rm RF}$ the number of RF chains available at the transmitter and at the receiver, respectively, the $k$-th user precoding and combining matrices are decomposed as follows:
\begin{equation}
\mathbf{Q}_k^{\rm PZF-HY}=\mathbf{Q}_k^{\rm RF} \mathbf{Q}_k^{\rm BB}\; , \quad
\mathbf{D}_k^{\rm PZF-HY}=\mathbf{D}_k^{\rm RF} \mathbf{D}_k^{\rm BB} \; .
\end{equation}
In the above decomposition, the matrices $\mathbf{Q}_k^{\rm RF}$ and $\mathbf{D}_k^{\rm RF}$ have dimension 
$(N_T \times N_T^{\rm RF})$ and $(N_R \times N_R^{\rm RF})$, respectively, and their entries are constrained to have constant (unit) norm (i.e. they are implemented through a network of phase-shifters\footnote{The case of quantized phase-shifts is also considered in the literature, but we are neglecting it here for the sake of simplicity.}); the matrices $\mathbf{Q}_k^{\rm BB}$ and $\mathbf{D}_k^{\rm BB}$, instead, have dimension $(N_T^{\rm RF} \times M)$ and $(N_R^{\rm RF} \times M)$, respectively, and their entries are unconstrained complex numbers. A block-scheme of the architecture of the hybrid transceiver is reported in Fig. \ref{Fig:hybrid_structure}.
Now, designing an hybrid beamformer is tantamount to finding expressions for the matrices $\mathbf{Q}_k^{\rm RF}, \mathbf{Q}_k^{\rm BB}, \mathbf{D}_k^{\rm RF},$ and $\mathbf{D}_k^{\rm BB}$, so that some desired beamformers are approximated. For the PZF-HY beamforming, the desired beamformers are the PZF-FD matrices, and their approximation is realized by using  the  block coordinate descent for subspace decomposition algorithm \cite{ghauch2015subspace}. We will assume that the number of RF chains in the BS is equal to $KM$, while at the mobile terminal it is equal to $M$. 

The amount of power consumed by the transmitter circuitry can be now written as \cite{MIMOArchitecture}:
\begin{equation}
P_{\rm{TX},c}=N_T^{\rm RF}\left(P_{\rm RFC}+P_{\rm DAC}+N_T P_{\rm PS}\right)+N_T P_{\rm PA}+P_{\rm BB} \;  ,
\end{equation} 
and the amount of power consumed by the receiver circuitry can be expressed as:
\begin{equation}
P_{\rm{RX},c}=N_R^{\rm RF}\left(P_{\rm RFC}+P_{\rm ADC}+N_R P_{\rm PS}\right)+N_T P_{\rm LNA}+P_{\rm BB} \;  .
\end{equation}
We have already given numerical values for the above quantities, except that for $P_{\rm PS}$, the power consumed by each  phase shifters, that we assume to be 30 mW as in \cite{MIMOArchitecture}.

\subsection{Fully Analog (AN) beamforming}
Fully analog beamforming requires that the entries of the precoding and combining matrices have constant norm. Here, we introduce a further constraint and assume that the columns of the matrices $\mathbf{Q}_k$ and $\mathbf{D}_k$ are unit-norm beam-steering vectors expressed as in Eq. \eqref{eq:ULA}.  Focusing on the generic $k$-th user, the columns of the matrix $\mathbf{Q}_k^{\rm AN}$ are chosen as the array responses corresponding to the departure angles in the channel model \eqref{eq:channel1} associated to the $M$ dominant paths. A similar choice is made for $\mathbf{D}_k^{\rm AN}$, whose columns contain the array responses corresponding to the $M$ arrival angles associated to the $M$ dominant paths. In order to avoid self-interference, we have included a further constraint in the choice of the dominant paths to ensure that the angles of departure (arrivals) of the selected paths are spaced of at least 5 deg. The amount of power consumed by the transmitter circuitry can be written as:
\begin{equation}
P_{\rm{TX},c}=N_T^{\rm RF}\left(P_{\rm RFC}+N_T P_{\rm element}+P_{\rm DAC}\right) \;  ,
\end{equation} 
and the amount of power consumed by the receiver circuitry can be expressed as:
\begin{equation}
P_{\rm{RX},c}=N_R^{\rm RF}\left(P_{\rm RFC}+N_R P_{\rm element}+P_{\rm ADC}\right) \;  ,
\end{equation}
where $P_{\rm element}= 27$ mW \cite{Phased_Array60GHz} is the power consumed by each element of the phased array.

\subsection{Beamforming based on switches and fixed phase shitfers (SW+PHSH)}
\begin{figure}
\begin{subfigure}
  \centering
  \includegraphics[scale=0.42]{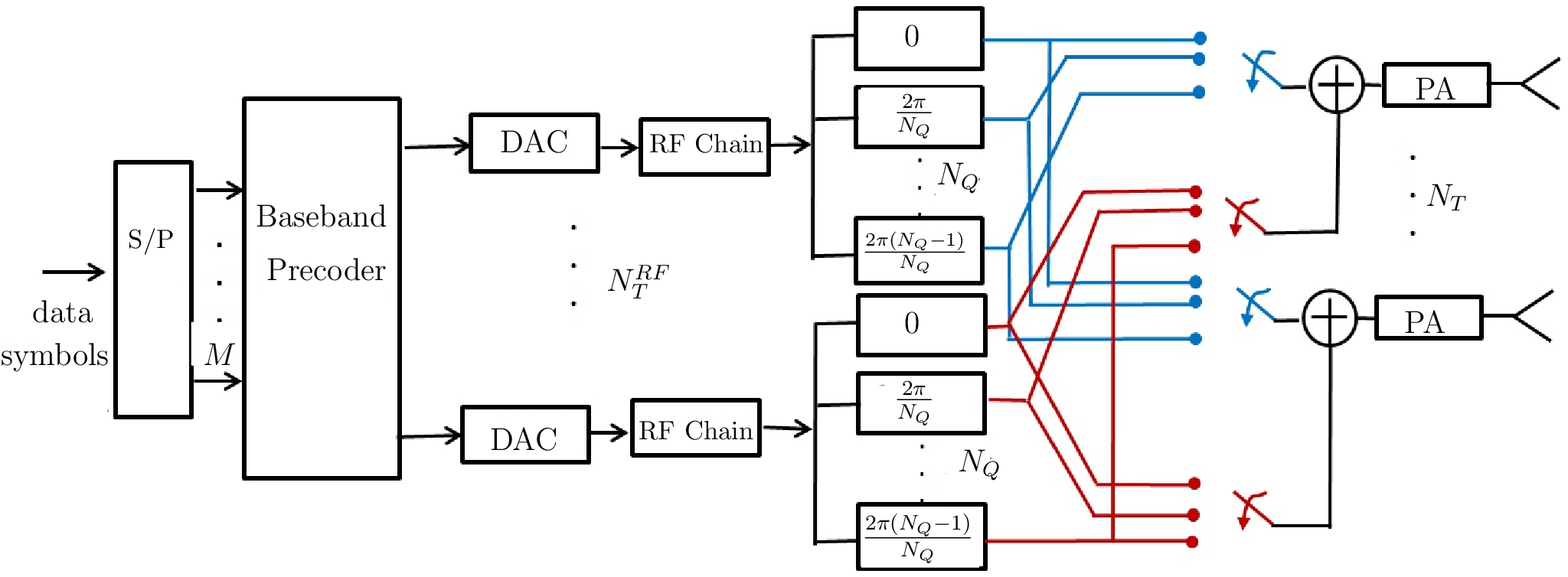}
\end{subfigure}%
\begin{subfigure}
  \centering
  \includegraphics[scale=0.42]{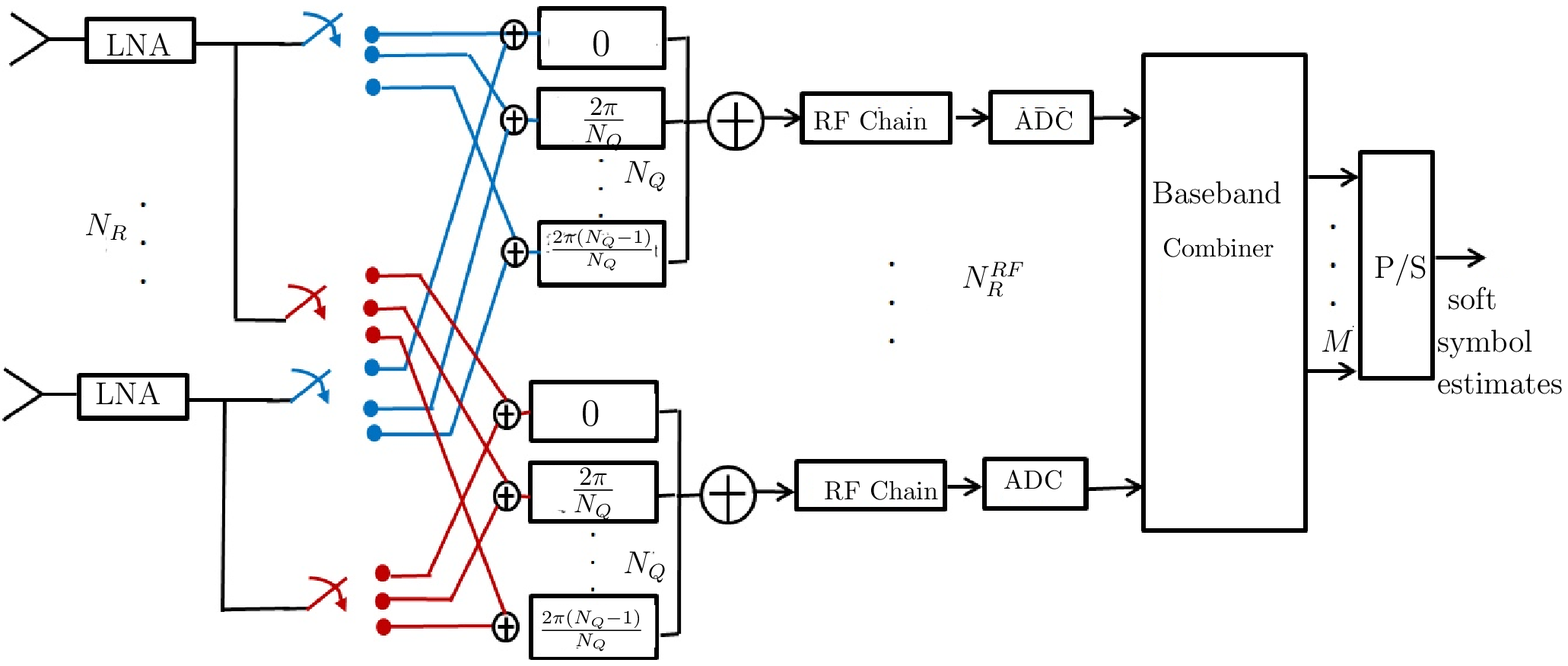}

\end{subfigure}  \label{fig:SW}
\caption{Block-scheme of a transceiver where beamforming is implemented with switches and $N_Q$ constant phase shifters per RF chain.}
\label{Fig:switchesPS_structure}
\end{figure}

The considered structure, depicted in Fig. \ref{Fig:switchesPS_structure},  builds upon the one reported in 
\cite{Switches_constantPS}, wherein a massive MIMO combiner is proposed based on the use of switches and fixed (i.e., not tunable) phase shifters. The scheme in Fig. \ref{Fig:switchesPS_structure} extends the structure of \cite{Switches_constantPS} by including also the precoding design. Due to lack of space, we omit providing all the details for the synthesis of the structure, and just explain the intuition behind the algorithm. The $(i,j)$-entry of the precoding matrix is in the form $[\mathbf{Q}^{\rm SW+PHSH}]_{(i,j)}=e^{\phi_{i,j}}$ , so it is an unitary module entry and its angle is obtained comparing the angle of the precoding matrix that we aim to synthesize 
with the quantization angles $\frac{2(q-1)\pi}{N_Q} \, , q=1, \ldots , N_Q$.  A similar reasoning is followed for the entries of the postcoding matrix $\mathbf{D}^{\rm SW+PHSH}$. We will assume that the number of quantization angles is $N_Q=8$ the number of RF chains in the BS is equal to $KM$, while at the mobile terminal it is equal to $M$. Both at the transmitter and the receiver we have $N_Q$ constant phase shifters per RF chain, and $N_T^{\rm RF}$ and $N_R^{\rm RF}$ switches per antenna at the transmitter and at the receiver, respectively. 

The amount of power consumed by the transmitter circuitry can be thus written as:
\begin{equation}
\begin{array}{lll}
P_{\rm{TX},c}= &N_T^{\rm RF}\left(P_{\rm RFC}+P_{\rm DAC}+N_Q P_{\rm PS}^{\rm fixed}\right)+ \\ & N_T\left(N_T^{\rm RF}P_{\rm SW}+P_{\rm PA}\right)+P_{\rm BB} \;  ,
\end{array}
\end{equation} 
and the amount of power consumed by the receiver circuitry can be expressed as:
\begin{equation}
\begin{array}{lll}
P_{\rm{RX},c}=&N_R^{\rm RF}\left(P_{\rm RFC}+P_{\rm ADC}+N_Q P_{\rm PS}^{\rm fixed}\right)+ \\ & N_R\left(N_R^{\rm RF}P_{\rm SW}+P_{\rm LNA}\right)+P_{\rm BB}  \;  .
\end{array}
\end{equation} 
In the above equations, $P_{\rm SW}=5$ mW \cite{MIMOArchitecture} is the power consumed by the single switch, and $P_{\rm PS}^{\rm fixed}$ is the power consumed by the constant phase shifter; this term  is of course lower than the power consumed by a tunable phase shifter, and we thus assume 1 mW.

\subsection{Switch-based (SW) beamforming}
A beamforming structure exclusively based on the use of switches is reported in  \cite{SwitchesRial}. We denote, again, with $N_T^{\rm RF}$ and $N_R^{\rm RF}$ the number of RF chains at the transmitter and at the receiver, respectively, and we assume that we have $N_T^{\rm RF}$ switches at the transmitter and $N_R^{\rm RF}$ at the receiver that select the antennas using the Minimum Frobenius Norm (MFN) algorithm in \cite{SwitchesRial}. The precoding matrix is in the form $\mathbf{Q}^{\rm SW}=\mathbf{S}\mathbf{Q^{\rm BB}}$ where $\mathbf{S}$ is a $N_T \times N_T^{\rm RF}$-dimensional matrix with columns that have exactly one position containing the value "1" and the other entries in matrix are zero, and $\mathbf{Q^{\rm BB}}$ is the $N_T^{\rm RF} \times M$-dimensional baseband precoding matrix. It can be thus shown that the matrix $\mathbf{Q}^{\rm SW}$ contains non-zero $N_T^{\rm RF}$ rows corresponding to the $N_T^{\rm RF}$ rows of the precoding matrix that we aim to synthesize with the largest norm.  A similar reasoning is followed for the entries of the postcoding matrix $\mathbf{D}^{\rm SW}$. We will, again, assume that the number of RF chains in the BS is equal to $KM$, while at the mobile terminal it is equal to $M$.

The amount of power consumed by the transmitter circuitry can be written as:
\begin{equation}
\begin{array}{lll}
P_{\rm{TX},c}=& N_T^{\rm RF}\left(P_{\rm RFC}+P_{\rm DAC}+P_{\rm SW}\right)+ 
 N_T^{\rm RF} P_{\rm PA}+P_{\rm BB} \;  ,
\end{array}
\end{equation} 
and the amount of power consumed by the receiver circuitry can be expressed as:
\begin{equation}
\begin{array}{lll}
P_{\rm{RX},c}= & N_R^{\rm RF}\left(P_{\rm RFC}+P_{\rm ADC}+P_{\rm SW}\right)+
N_R^{\rm RF} P_{\rm LNA}+P_{\rm BB}  \,  .
\end{array}
\end{equation}

\begin{figure}[t]
\includegraphics[scale=0.48]{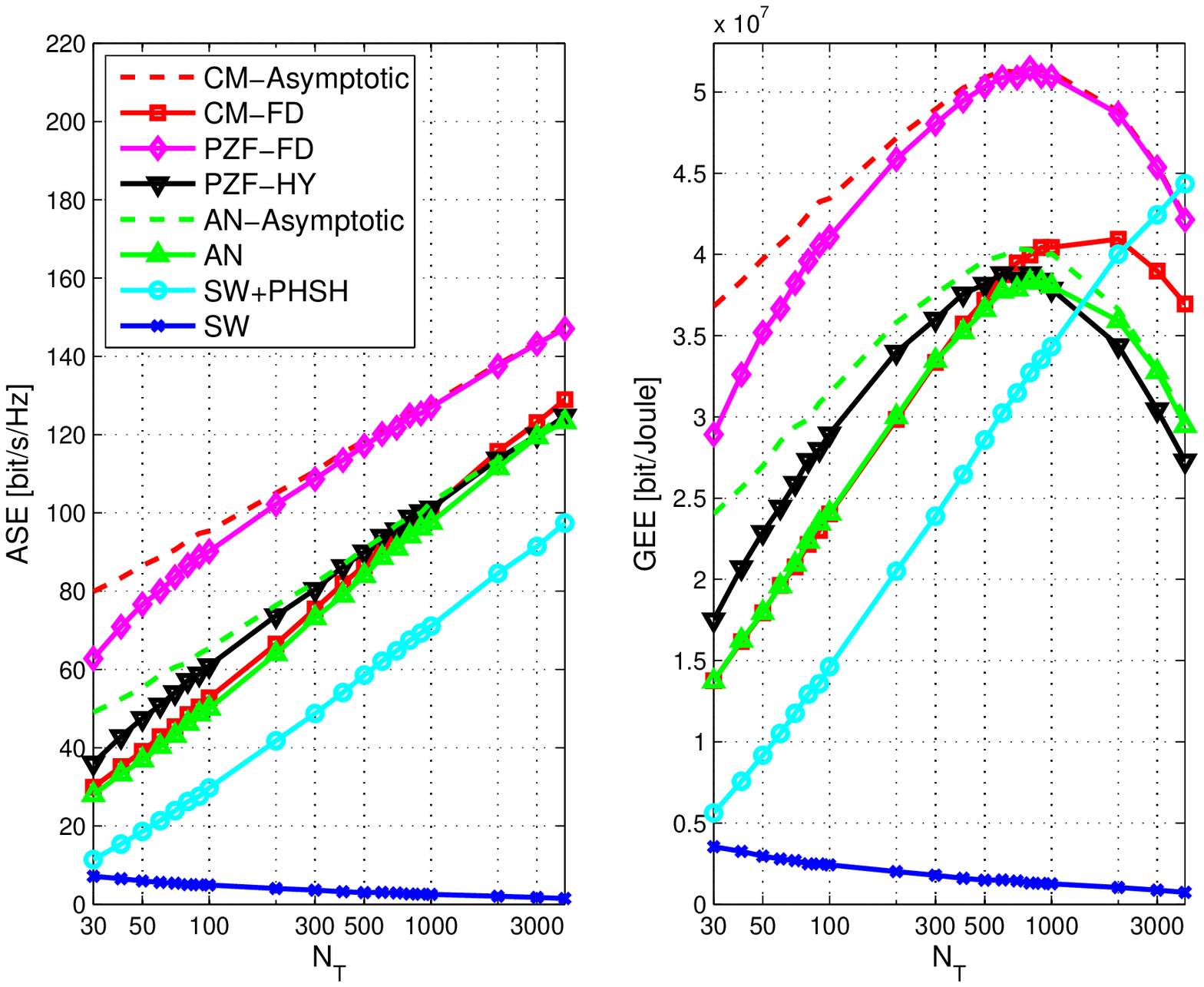}
\caption{Plot of ASE and GEE versus $N_T$ with $M=1$ and $P_T=0$ dBW.}
\label{Fig:1M1}
\end{figure}

\begin{figure}[t]
\includegraphics[scale=0.48]{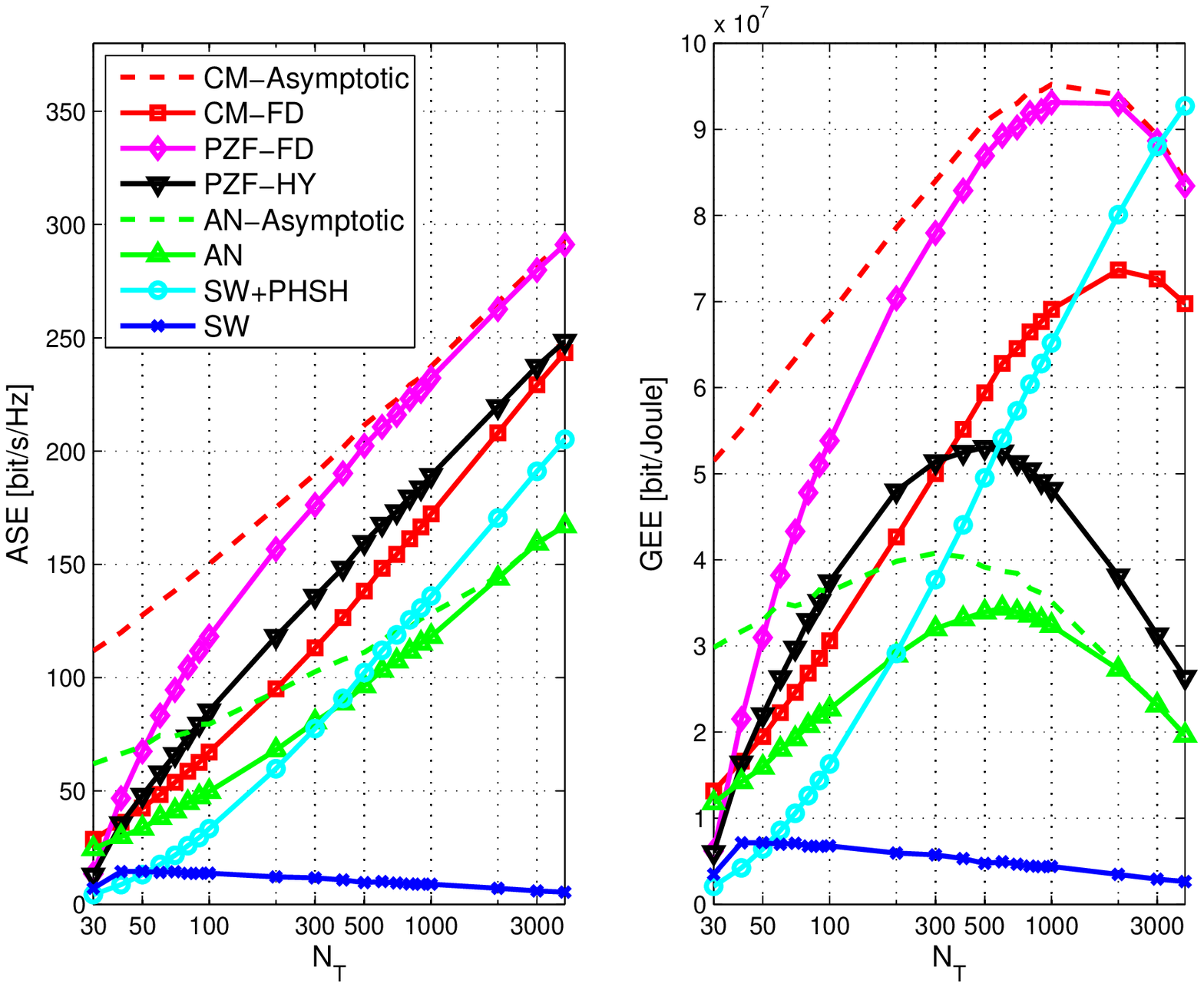}
\caption{Plot of ASE and GEE versus $N_T$ with $M=3$ and $P_T=0$ dBW.}
\label{Fig:2M3}
\end{figure}

\begin{figure}[t]
\includegraphics[scale=0.48]{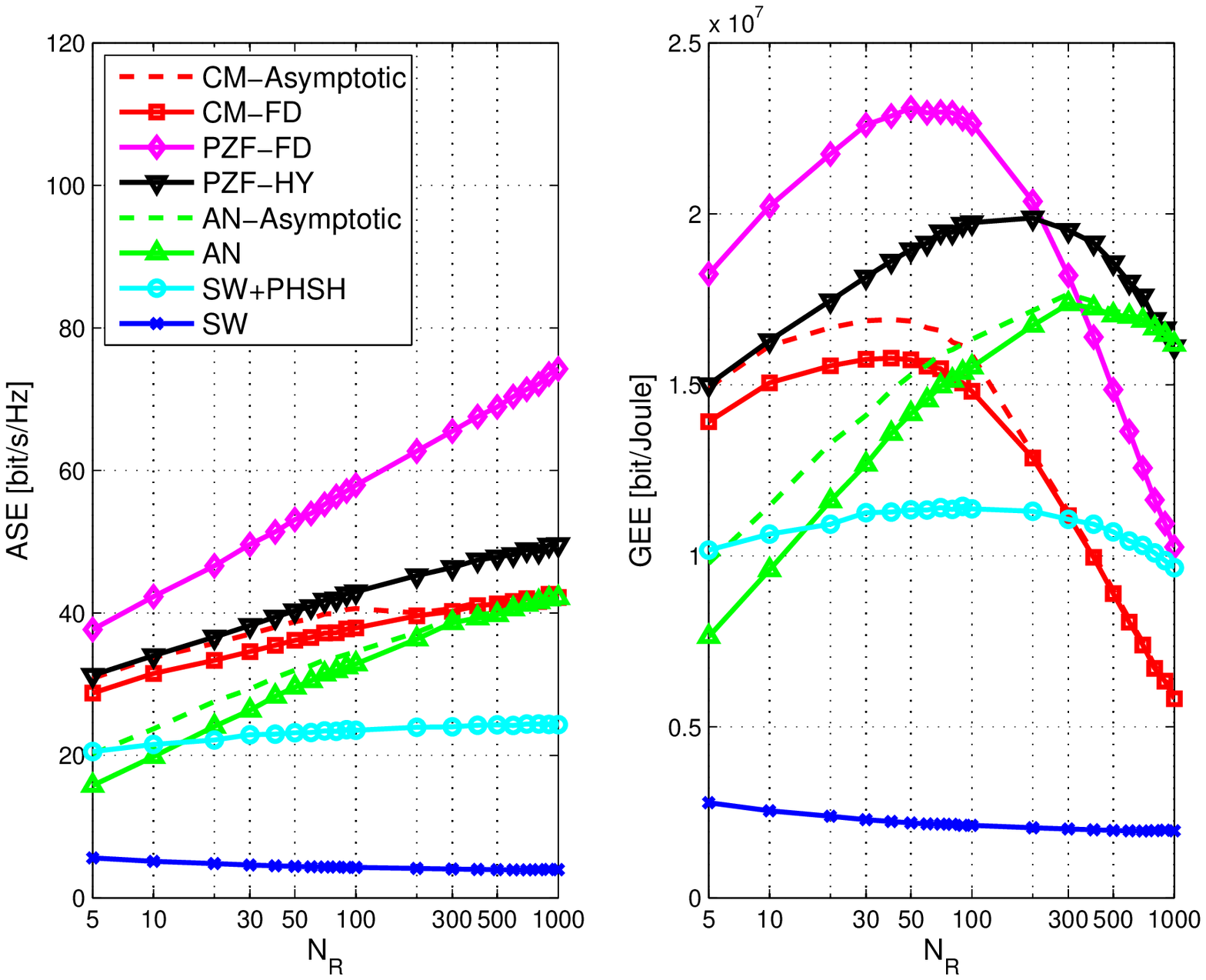}
\caption{Plot of ASE and GEE versus $N_R$ with $M=1$ and $P_T=0$ dBW.}
\label{Fig:3M1}
\end{figure}

\begin{figure}[t]
\includegraphics[scale=0.48]{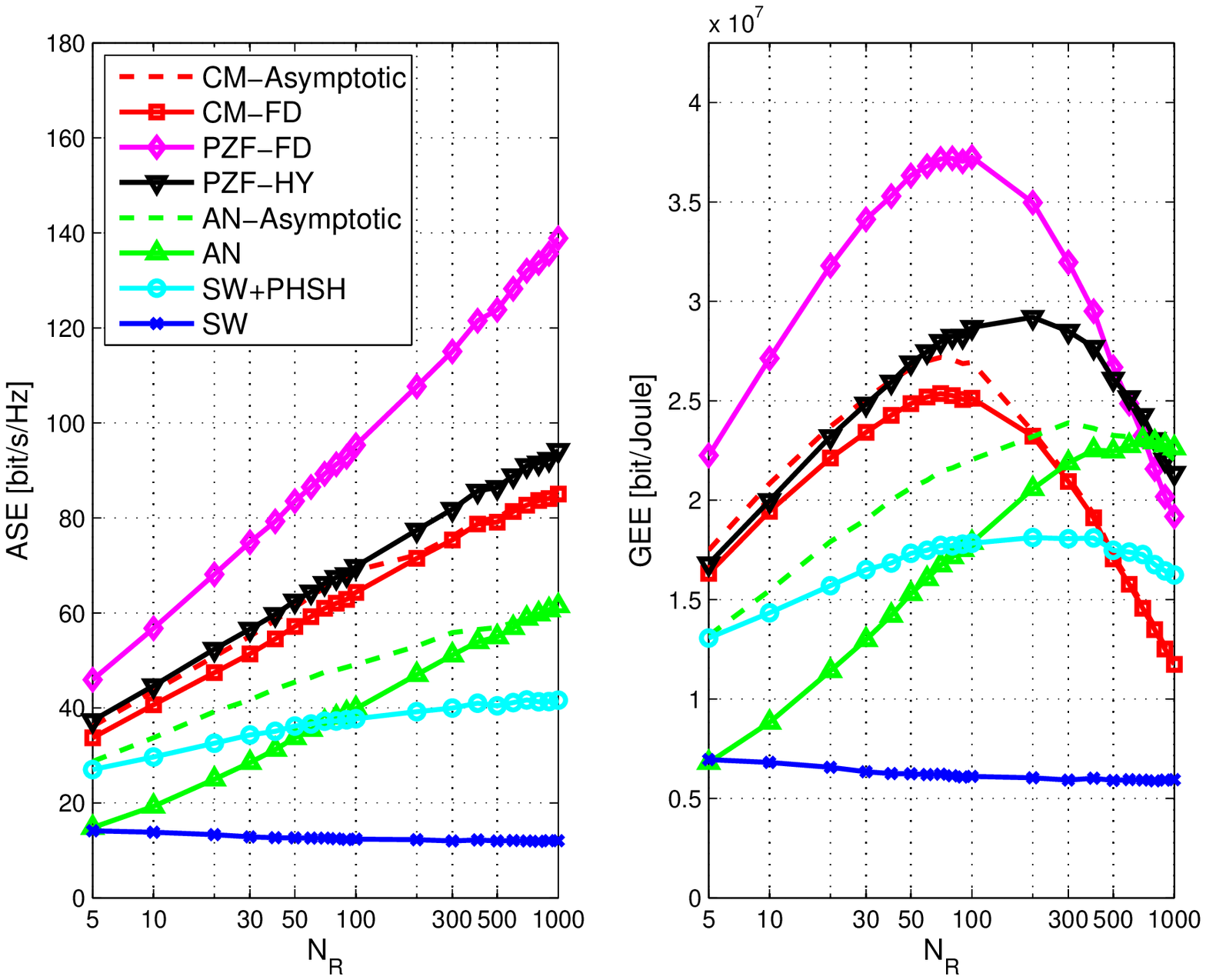}
\caption{Plot of ASE and GEE versus $N_R$ with $M=3$ and $P_T=0$ dBW.}
\label{Fig:4M3}
\end{figure}

\section{Numerical results}
We now provide simulation results showing the ASE and the GEE for the downlink of a single-cell mmWave system. We assume that there are $K=10$ users using the same frequency band and whose locations  are random, with $100$ m maximum distance from the BS. The parameters for the generation of the matrix channels are the ones reported in \cite{buzzidandreachannel_model}
for the ``street canyon model'', with $N_{\rm cl}=2$ and $N_{{\rm ray},i}=20, \; \forall i=1 \ldots
N_{\rm cl}$.
We assume that the BS transmit power is 0 dBW, the carrier frequency $f_c$=73 GHz, the used bandwidth is $W=500$ MHz,  the noise power $\sigma^2_n= F {\cal N}_0 W$, with $F=3$ dB the receiver noise figure and 
${\cal N}_0= -174$ dBm/Hz. 
In the following,  we report results as a function of the number of transmit and receive antennas. 
Figs. \ref{Fig:1M1} and \ref{Fig:2M3} report the ASE and the GEE versus the number of transmit antennas (assuming $N_R=30$) for the case in which $M=1$ and $M=3$, respectively. Figs. \ref{Fig:3M1}  and \ref{Fig:4M3}, similarly, show the ASE and the GEE versus the number of receive antennas (assuming $N_T=100$), again for $M=1$ and $M=3$, respectively. Inspecting the figures, it is seen that the best performing beamforming structure is the PZF-FD, both in terms of ASE and of GEE. This last conclusion is quite surprising, since it shows that lower complexity structures, although necessary for obvious practical considerations, actually are less energy efficient (from a communication physical layer perspective) than fully digital structures. Results also show that the SW structure achieves quite unsatisfactory performance, while, on the contrary, the fully analog (AN) structure performs, at least in the case that $M=1$, approximately as the CM-FD structure. We can also note that, for low values of $N_T$ (and of $N_R$ too), the PZF-FD and its hybrid approximation PZF-HY are the best solutions. Conversely, for very large values of $N_T$, the solution with the largest GEE is ths SW+PHSH one; for large values of $N_R$, instead, the best-GEE solution is the AN beamforming structure. Figs. \ref{Fig:1M1} -- \ref{Fig:4M3} also report, as dashed lines, the behavior of asymptotic analytical expressions\footnote{These asymptotic formulas are not given here for the sake of brevity and will be reported elsewhere.} for the AN structure and for the CM-FD structure; we note that the asymptotic AN formula tightly approaches the performance of the AN structure, for large number of antennas, whereas, the asymptotic CM formula approaches the performance of the CM-FD structure only for large values of $N_R$, while for large values of $N_T$ the performamce of the PZF-FD structure is approached.

\section{Conclusions}
In this paper we have provided an analysis of both the spectral efficiency and the energy efficiency of a downlink MU-MIMO system operating at mmWave frequencies and with several fully digital and low-complexity beamforming architectures. Our results have revealed that, using some of the most recent available data on the energy consumption of transceiver components, fully-digital architectures, although unfeasible for large number of antennas due to complexity constraints, are superior not only in terms of achievable rate (as it was largely expected), but also in terms of energy efficiency. In particular, among fully-digital implementations, the PZF-FD architecture has been shown to provide the best performance, while, among the lower complexity implementations, the most relevant alternatives are the AN and the SW+PHSH structures. Of course the provided results and the relative ranking among the considered structures is likely to change in the future as technology progresses and devices with reduced power consumption appear on the scene. 
The analysis provided in this paper has assumed a uniform power splitting among users and data-streams; it is expected that improved performance can be obtained through waterfilling-like power control. Moreover, we have assumed Gaussian-distributed data, while, instead, the effect of finite-cardinality modulation is also worth being investigated. Finally, the provided results have been derived under the assumption of perfect channel state information, and it is thus of interest to extend this work to the case in which there are channel estimation errors.

\bibliographystyle{IEEEtran}

\bibliography{FracProg_SB,finalRefs,references}

%



\end{document}